\documentclass[aps,pra,superscriptaddress,twocolumn,showpacs]{revtex4}
\usepackage{amsfonts}
\usepackage{amsmath}
\usepackage{amssymb}
\usepackage{graphicx}

\begin{document}

\title{Structure of two-component Bose-Einstein condensates with respective
vortex-antivortex superposition states}
\author{Linghua Wen}
\email{linghuawen@126.com}
\affiliation{School of Physical Sciences and Information Engineering, Liaocheng
University, Liaocheng 252059, China}
\affiliation{Department of Physics, The University of Texas at Dallas, Richardson, Texas
75080, USA}
\author{Yongjun Qiao}
\affiliation{School of Physical Sciences and Information Engineering, Liaocheng
University, Liaocheng 252059, China}
\author{Yong Xu}
\affiliation{Institute of Physics, Chinese Academy of Sciences, Beijing 100190, China}
\affiliation{Department of Physics, The University of Texas at Dallas, Richardson, Texas
75080, USA}
\author{Li Mao}
\affiliation{Department of Physics, The University of Texas at Dallas, Richardson, Texas
75080, USA}
\date{\today }

\begin{abstract}
We investigate the phase structure of two-component Bose-Einstein
condensates (BECs) with repulsive intra- and interspecies interactions in
the presence of respective vortex-antivortex superposition states (VAVSS).
We show that different winding numbers of vortex and antivortex and
different intra- and interspecies interaction strengths may lead to
different phase configurations, such as fully separated phases, inlaid
separated phases, asymmetric separated phase, and partially mixed phases,
where the density profile of each component displays a petal-like (or
modulated petal-like) structure. A phase diagram is given for the case of
equal unit winding numbers of the vortex and antivortex in respective
components, and it is shown that conventional criterion for phase separation
of two-component BECs is not applicable for the present system due to the
VAVSS. In addition, our nonlinear stability analysis indicates that the
typical phase structures of two-component BECs with VAVSS allow to be
detected in experiments. Moreover, for the case of unequal winding numbers
of the vortex and antivortex in respective components, we find that each
component in any of the possible phase structures is in a cluster state of
vortices and antivortices, where the topological defects appear in the form
of singly quantized visible vortex, or hidden vortex, or ghost vortex,
depending on the specific parameters of the system. Finally, a general rule
between the vortex-antivortex cluster state and the winding numbers of
vortex and antivortex is revealed.
\end{abstract}

\maketitle

\section{Introduction}

Multicomponent systems are of fundamental importance in many fields of
physics on account of their universality and complexity. In particular,
great attention has been recently paid to a mixture of two-component
Bose-Einstein condensates (BECs) in cold atom physics due to ultra high
purity, excellent theoretical description, experimental accessibility and
precise controllability of a condensate system \cite{Pethick,Pitaevskii}.
Two-component BECs can consist of two different alkalis, or different
isotopes, or different hyperfine states of the same alkali atom. Such a
mixture provides an attractive and versatile testing ground to study the
intriguing properties of macroscopic quantum many-body systems inaccessible
in single-component BECs. In fact, many interesting phenomena have been
predicted theoretically and some observed experimentally in two-species
BECs, ranging from the topological structures of the ground and excited
states \cite%
{Ho,Pu,Trippenbach,Adhikari1,Riboli,Navarro,Catelani,Gligoric,LWen,Kuopanportti,Papp,McCarron}%
, symmetry-breaking transition \cite{Lee}, quantum turbulence \cite%
{Takeuchi1}, pattern formation \cite{Sabbatini}, vortex-bright-solitons \cite%
{Law} and skyrmions \cite{Kawakami}, entangled states \cite{Micheli}, to the
stripe phase induced by spin-orbit coupling \cite{Zhang}, etc. Among these
phenomena, of crucial significance is the phase structure of the system \cite%
{Ho,Pu,Trippenbach,Adhikari1,Riboli,Navarro,Catelani,Gligoric,LWen,Kuopanportti,Papp,McCarron}%
. For instance, a separated phase of a two-component BEC is the premise of
studying the Kelvin-Helmholtz instability \cite{Takeuchi2} and
Rayleigh-Taylor instability \cite{Sasaki}.

To the best of our knowledge, most of theoretical literatures use the
Thomas-Fermi (TF) approximation (a semi-analytical method) or a variational
approach to study the ground state properties of two-component BECs, where
spherically symmetric mixed phases, spherically symmetric separated phases,
and asymmetric side-by-side separated phases are predicted. For the former
case, the kinetic energy is completely neglected. However, when the
contribution of kinetic energy to the total energy is considerable, the TF
approximation will fail \cite{Catelani} and thus can not be relied upon to
determine the phase structure of a two-component BEC. The effect of kinetic
energy on the phase transition of a two-component BEC in an infinitely deep
square well potential has been recently demonstrated in Ref. \cite{LWen}. As
for the variational approach, it is usually limited to few simple quantum
states and certain external potentials. In this context, it is necessary to
resort to numerical methods for obtaining reliable phase structures.

In this paper, we investigate the exact two-dimensional (2D) phase
structures of two-component BECs with respective vortex-antivortex
superposition states (VAVSS). The VAVSS in BECs is a quite interesting
research object as it can exhibit rich physical properties \cite%
{Kapale,Liu,Simula,Thanvanthri,Wen1}, such as peculiar petal-like structure
\cite{Kapale,Liu} and unique dynamics \cite{Wen1}. Particularly, the
equilibrium properties of a quasi-2D degenerate boson--fermion mixture
(DBFM) with a bosonic VAVSS is recently studied by using a
quantum-hydrodynamic model \cite{Wen2}. It is shown that the VAVSS greatly
influences the equilibrium state and stability of a DBFM, where a special
intermittency phenomenon\ exists in the two stability curves of the DBFM
with a bosonic VAVSS \cite{Wen2} and the one with a bosonic vortex \cite%
{Adhikari2}. On the other hand, the VAVSS may have many potential
applications in quantum information, quantum communication, and inertial
sensing \cite{Kapale,Simula,Thanvanthri}. Experimentally, the creation of
VAVSS in BECs has been reported by different groups \cite%
{Andersen,Wright1,Wright2,Scherer}. An interesting question is to ask how
the phase structure of a two-component BEC is modified by the respective
VAVSS combining with intra- and interspecies interactions. We show that,
depending on the relative interaction strengths and the winding numbers of
vortex and antivortex, the two-component BECs with respective VAVSS can
display rich phase structures not met in other systems, such as fully
separated phases, inlaid separated phases, and asymmetric separated phase
with (deformed) petal-like component density profiles. We give a phase
diagram for the case of equal unit winding numbers of vortex and antivortex
in respective components. The typical phase structures are long-lived
according to our nonlinear stability analysis. For the case of unequal
winding numbers of vortex and antivortex, each component in any of the
possible phases is in a vortex-antivortex cluster state, where the phase
defects emerge in the form of visible vortex, or hidden vortex, or ghost
vortex. Furthermore, we reveal and discuss the general relation between the
vortex-antivortex cluster state and the winding numbers of vortex and
antivortex.

The paper is organized as follows. In Sec. II, we describe the model for
two-component BECs with respective VAVSS. In Sec. III, we study the phase
structure of two-component BECs with respective VAVSS with equal unit
winding numbers of the vortex and antivortex. A phase diagram is presented,
and the stability of phase structures is analyzed. In Sec. IV, we discuss
the phase structure of the system with unequal winding numbers of the vortex
and antivortex in respective components. The conclusion is outlined in the
last section.

\section{Model}

We consider a 2D system of trapped two-component BECs with tight confinement
in the $z$ direction. In the zero-temperature limit, the dynamics of the
system is described by the coupled Gross-Pitaevskii (GP) equations
\begin{eqnarray}
i\frac{\partial \psi _{1}}{\partial t} &=&\left[ -\nabla _{\perp
}^{2}+V(r)+g_{1}\left\vert \psi _{1}\right\vert ^{2}+g\left\vert \psi
_{2}\right\vert ^{2}\right] \psi _{1},  \label{GPE1} \\
i\frac{\partial \psi _{2}}{\partial t} &=&\left[ -\nabla _{\perp
}^{2}+V(r)+g_{2}\left\vert \psi _{2}\right\vert ^{2}+g\left\vert \psi
_{1}\right\vert ^{2}\right] \psi _{2},  \label{GPE2}
\end{eqnarray}%
where we have adopted plane-polar coordinates for space $(r,\theta )$, $%
\nabla _{\perp }^{2}=\frac{1}{r}\frac{\partial }{\partial r}(r\frac{\partial
}{\partial r})+\frac{1}{r^{2}}\frac{\partial ^{2}}{\partial \theta ^{2}}$,
and $V(r)=r^{2}/4$ is the trapping potential. Here we have assumed that the
two species have the same atomic mass $m$ and undergo the same trapping
frequency $\omega _{\bot }$ in the $x$-$y$ plane. In this paper, length,
time, energy, and angular momentum are in units of $d_{0}=\sqrt{\hbar
/2m\omega _{\bot }}$, $1/\omega _{\bot }$, $\hbar \omega _{\bot }$, and $%
\hbar $, respectively. $g_{1},g_{2},$and $g$ denote dimensionless intra- and
intercomponent coupling strengths which are proportional to the
corresponding $s$-wave scattering lengths $a_{j},a_{ij}$ ($j=1,2$) between
intra- and intercomponent atoms and the atom numbers $N_{1,2}$ in species $1$
and $2$ \cite{Pethick,Pitaevskii,Wen1,Wen2}. The two wave functions are
normalized as $\iint \left\vert \psi _{1,2}\right\vert ^{2}rdrd\theta =1$.

We seek dimensionless solutions of stationary VAVSS%
\begin{equation}
\psi _{j}(r,\theta ,t)=\varphi _{j}(r,\theta )e^{-i\mu _{j}t},\ \ j=1,2,
\label{TrialWFCartesian}
\end{equation}%
where $\mu _{j}$ is the chemical potential, and $\varphi _{j}$ is given by%
\begin{equation}
\varphi _{j}(r,\theta )=f_{jk}\alpha _{j}e^{il_{jk}\theta }+f_{jp}\beta
_{j}e^{i\delta _{j}}e^{-il_{jp}\theta },\text{\ \ }j=1,2.
\label{PolarTimeInWF}
\end{equation}%
Here $f_{jk}(r)=A_{j}e^{-r^{2}/2\sigma _{j}^{2}}(r/\sigma _{j})^{l_{jk}}$
and $f_{jp}(r)=A_{j}e^{-r^{2}/2\sigma _{j}^{2}}(r/\sigma _{j})^{l_{jp}}$ are
the amplitudes of the vortex and antivortex components with $A_{j}$ being
the normalization constant and $\sigma _{j}$ the width of condensate $j$. $%
l_{jk\text{ }}$and $l_{jp}$ are the winding numbers of the vortex and
antivortex, and the real constants $\alpha _{j}$, $\beta _{j}$ show the
proportion of the vortex and antivortex with $\alpha _{j}^{2}+\beta
_{j}^{2}=1$. The relative phase $\delta _{j}$ just causes offset of the
density profile by an angle $\delta _{j}/(l_{jk\text{ }}+l_{jp})$. Note that
in Eq. (\ref{PolarTimeInWF}) we have taken into account the influence of
winding number on the condensate size in terms of $\sigma _{j}^{-l_{jk}}$
and $\sigma _{j}^{-l_{jp}}$.

Substituting Eq. (\ref{TrialWFCartesian}) into Eqs. (\ref{GPE1}) and (\ref%
{GPE2}), we obtain the coupled equations for $\varphi _{1}$ and $\varphi
_{2} $,
\begin{eqnarray}
\mu _{1}\varphi _{1} &=&-\frac{1}{r}\frac{\partial }{\partial r}(r\frac{%
\partial \varphi _{1}}{\partial r})+\frac{f_{1p}\beta _{1}}{r^{2}}%
(l_{1p}^{2}-l_{1k}^{2})e^{i(\delta _{1}-l_{1p}\theta )}  \notag \\
&&+\frac{l_{1k}^{2}}{r^{2}}\varphi _{1}+(V+g_{1}\left\vert \varphi
_{1}\right\vert ^{2}+g\left\vert \varphi _{2}\right\vert ^{2})\varphi _{1},
\label{PolarTimeInd1} \\
\mu _{2}\varphi _{2} &=&-\frac{1}{r}\frac{\partial }{\partial r}(r\frac{%
\partial \varphi _{2}}{\partial r})+\frac{f_{2p}\beta _{2}}{r^{2}}%
(l_{2p}^{2}-l_{2k}^{2})e^{i(\delta _{2}-l_{2p}\theta )}  \notag \\
&&+\frac{l_{2k}^{2}}{r^{2}}\varphi _{2}+(V+g_{2}\left\vert \varphi
_{2}\right\vert ^{2}+g\left\vert \varphi _{1}\right\vert ^{2})\varphi _{2},
\label{PolarTimeInd2}
\end{eqnarray}%
where the second and third terms associated with winding numbers in the
right-hand sides of Eqs. (\ref{PolarTimeInd1}) and (\ref{PolarTimeInd2}) are
resulted from the presence of VAVSS in each component. Consequently, the
chemical potentials $\mu _{1,2}$ read%
\begin{eqnarray}
\mu _{1} &=&\frac{2(l_{1k}+1)}{\sigma _{1}^{2}}-\frac{1}{\sigma _{1}^{4}}%
\iint \left\vert \varphi _{1}\right\vert ^{2}r^{3}drd\theta  \notag \\
&&+\frac{4\pi (l_{1p}-l_{1k})}{\sigma _{1}^{2}}\int f_{1p}^{2}\beta
_{1}^{2}rdr  \notag \\
&&+\iint [V+g_{1}\left\vert \varphi _{1}\right\vert ^{2}+g\left\vert \varphi
_{2}\right\vert ^{2}]\left\vert \varphi _{1}\right\vert ^{2}rdrd\theta ,
\label{ChemPotential1} \\
\mu _{2} &=&\frac{2(l_{2k}+1)}{\sigma _{2}^{2}}-\frac{1}{\sigma _{2}^{4}}%
\iint \left\vert \varphi _{2}\right\vert ^{2}r^{3}drd\theta  \notag \\
&&+\frac{4\pi (l_{2p}-l_{2k})}{\sigma _{2}^{2}}\int f_{2p}^{2}\beta
_{2}^{2}rdr  \notag \\
&&+\iint [V+g_{2}\left\vert \varphi _{2}\right\vert ^{2}+g\left\vert \varphi
_{1}\right\vert ^{2}]\left\vert \varphi _{2}\right\vert ^{2}rdrd\theta ,
\label{ChemPotential2}
\end{eqnarray}%
and they must be determined self-consistently from the normalization
conditions of $\varphi _{1}$ and $\varphi _{2}$.

From Eqs. (\ref{PolarTimeInd1})-(\ref{ChemPotential2}) we can see that when
the kinetic energy is comparable with the sum of interaction energy and
potential energy the TF approximation is not suitable to describe the
system, especially for cases of VAVSS with large winding numbers. In the
following, we numerically solve the coupled equations (\ref{PolarTimeInd1})
and (\ref{PolarTimeInd2}). Starting with two trial VAVSS with specific
ratios of $\alpha _{j}^{2}$ and $\beta _{j}^{2}$, we obtain the exact 2D
equilibrium state of the system by using the imaginary time propagation
method based on the Peaceman-Rachford method \cite{Peaceman}, which is
equivalent to the procedure of minimizing the sum of the average energy per
atom in species $1$ and that in species $2$,%
\begin{eqnarray}
E &=&\iint rdrd\theta \lbrack \mu _{1}\left\vert \varphi _{1}\right\vert
^{2}+\mu _{2}\left\vert \varphi _{2}\right\vert ^{2}-\frac{1}{2}%
g_{1}\left\vert \varphi _{1}\right\vert ^{4}  \notag \\
&&-\frac{1}{2}g_{2}\left\vert \varphi _{2}\right\vert ^{4}-g\left\vert
\varphi _{1}\right\vert ^{2}\left\vert \varphi _{2}\right\vert ^{2}].
\label{Energy}
\end{eqnarray}

\section{Structure of two-component BECs with respective VAVSS with equal
winding numbers of vortex and antivortex}

Here we just consider the case of $\alpha _{j}^{2}=\beta _{j}^{2}=1/2$
because the petal structure of a VAVSS with unequal superposition ratio only
emerges under the condition of a very small particle number or an extremely
weak interatomic interaction which is usually not met in experiments \cite%
{Wen2}. For convenience, we introduce two relative interaction strengths, $%
R_{21}=g_{2}/g_{1}$ and $R=g/g_{1}$, and assume the intra- and interspecies
interactions to be repulsive. Throughout this paper the relative phase is
taken to be $\delta _{j}=0$. Figure 1 shows the phase diagram of two BECs
with VAVSS with $l_{1k}=l_{1p}=1$ and $l_{2k}=l_{2p}=1$, where $g_{1}=200$.
There exist eight possible phases depending on the values of $R_{21}$ and $R$%
, and the typical density profiles corresponding to phases I-VIII are
displayed in Fig. 2 and Figs. 3(a)-3(b), respectively. In any of these
phases, the density profile of each component displays a petal structure or
a crescent-pair (deformed petal) structure or a combined structure of petal
and crescent-pair due to the presence of respective VAVSS and the
competition between the intraspecies and the interspecies repulsions. In
Fig. 1, regions I and IV denote two opposite fully separated phases. In the
former phase, component $1$ is completely expelled outside component $2$
[see Figs. 2(a1) and 2(a2)], while in the latter phase the sequence is
converse [Figs. 2(d1) and 2(d2)]. Regions II and III represent two opposite
inlaid separated phases, where in the former case species $1$ lies
separately in the interval between the inner petal layer and the outer
crescent layer of species $2$, however in the latter case the arrangement of
the two species is inverted [Figs. 2(b1)-2(c2)].\ Regions V and VI are two
different partially mixed phases, where in the first phase species $2$ is
merged inside species $1$ but it is reversed in the second phase [Figs.
2(e1)-2(f2)]. Finally, the vertical solid-line section ($0\leq R\leq 3.5$)
of $R_{21}=1$ marks a fully mixed phase in which the mixing reaches the
maximum and the densities profiles of two components are the same [Figs.
3(a1) and 3(a2)], while the vertical dashed-line section ($R>3.5$) signs an
asymmetric separated phase in which both the density profiles are symmetric
breaking upon the trap center [Figs. 3(b1) and 3(b2)].

\begin{figure}[tbp]
\centerline{\includegraphics*[width=6cm]{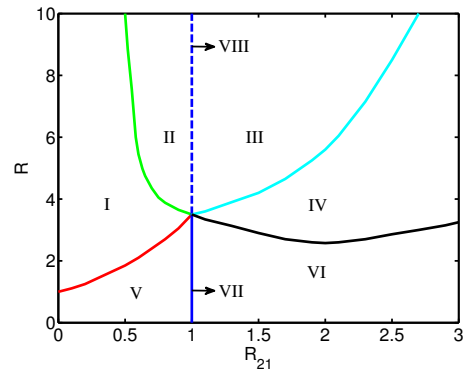}}
\caption{(color online) Phase diagram of two-component BECs with VAVSS with $%
l_{jk}=l_{jp}=1$ $(j=1,2)$, where $R_{21}=g_{2}/g_{1}$ and $R=g/g_{1}$. The
parameters are $g_{1}=200$ and $\protect\alpha _{j}^{2}=\protect\beta %
_{j}^{2}=1/2$. Regions I and IV, II and III, and V and VI represent two
different fully separated phases, inlaid separated phases, and partially
miscible phases, respectively. Vertical solid-line section VII marks a fully
mixed phase, while dashed-line section VIII denotes an asymmetric separated
phase.}
\label{Figure1}
\end{figure}

The ground state structure of two-component BECs has been studied by several
theoretical \cite{Ho,Pu,Trippenbach,Adhikari1,Riboli,Navarro,Gligoric,LWen}
and experimental groups \cite{Papp,McCarron}. A conventional criterion for
phase separation is given by $g_{1}g_{2}<g^{2}$, which is based on the
minimization of the total interaction energy \cite{Pethick,Pitaevskii}. Note
that in the derivation of this criterion the BECs are supposed to be
homogeneous and the kinetic energy is neglected. However, the criterion will
fail when the kinetic energy becomes important in a nonuniform system. In
fact, the kinetic energy play a vital role in determining the configuration
of two-component BECs with VAVSS. According to the condition $%
g_{1}g_{2}<g^{2}$, for instance, our system would be in a separated phase if
$R_{21}=2$ and $R>\sqrt{2}$, but only $R>2.58$ can the actual phase
separation occur when $R_{21}=2$. The similar effect arises for any fixed
value of $R_{21}$ or $R$ as shown in Fig. 1. Physically, the kinetic energy
acts against the interspecies interaction. The latter is responsible for
phase demixing while the former tends to expand the BECs and thus favors
phase mixing. At the same time, the trapping potential tends to trap the
condensates and hence also sustains phase mixing. Therefore, phase
separation can be suppressed by the kinetic energy and external potential in
some situations even if the condition $g_{1}g_{2}<g^{2}$ is satisfied, as we
saw in Figs. 1-3. Recently, the influence of kinetic energy on the
mixing-demixing transition by changing the confinement was discussed in Ref.
\cite{LWen}. The authors introduce a parameter $\eta =\int d\overrightarrow{r%
}\psi _{1}\psi _{2}$ to characterize the overlap between two condensate wave
functions. And then the system shows phase separation if $\eta \ll 1$.
However, neither the criterion $g_{1}g_{2}<g^{2}$ nor the condition $\eta
\ll 1$ can distinguish different separated phases as shown in Fig. 2 and
Fig. 3(b). In addition, from Eqs. (\ref{GPE1}), (\ref{GPE2}), (\ref%
{PolarTimeInd1}) and (\ref{PolarTimeInd2}), we can see that the two
components of the system satisfy the exchange symmetry, which indicates that
the phase diagram will be the same if one exchanges the two component wave
functions. For fixed values of $g_{1}$ and $g$, nevertheless, two different
values of $g_{2}$ which are symmetric concerning the line of $R_{21}=1$ will
lead to two different quantum states. The corresponding symmetric one of the
interaction set $(g_{1},g_{2},g)$ is $(g_{2},g_{1},g)$ [$g_{1}$ swaps $g_{2}$%
, and the set $(g_{2},g_{1},g)$ is not in the same phase diagram] rather
than the set $(g_{1},2g_{1}-g_{2},g)$ being symmetric about the line of $%
R_{21}=1$. This point can explain why the phase diagram in Fig. 1 is
asymmetric with respect to the line of $R_{21}=1$. Here $g_{1}=200$ is a
typical parameter value in current experiments. If the value of $g_{1}$ does
not vary largely, the phase diagram will be basically unchanged except for
possible little offset of the boundary lines between different phases. When
the value of $g_{1}$ changes largely, there exist similar phase structures
and the regions corresponding to different phase structures will possibly
redistribute in the phase diagram.

\begin{figure}[tbp]
\centerline{\includegraphics*[width=7.6cm]{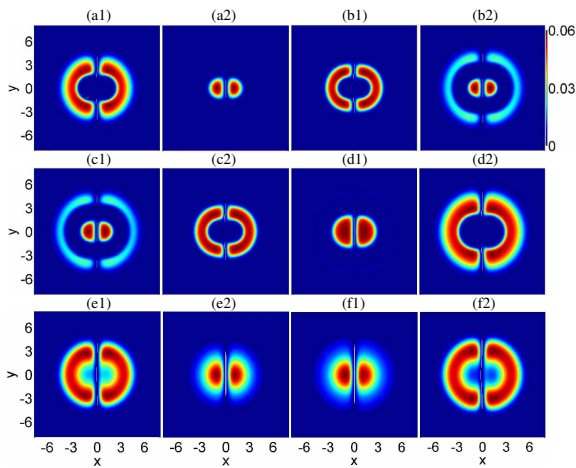}}
\caption{(color online) Density profiles (a)-(f) correspond to phases I-VI
in Fig.1, respectively. Here $1$ and $2$ label two different components of
the system. (a) $R_{21}=0.2,R=8,$(b) $R_{21}=0.8,R=8,$(c) $R_{21}=1.5,R=7,$%
(d) $R_{21}=2.5,R=5,$(e) $R_{21}=0.6,R=1,$and (f) $R_{21}=1.4,R=1.6$. The
other parameters are the same as those in Fig.1. The darker color area
indicates the lower density. $x$ and $y$ are in units of $d_{0}$.}
\label{Figure2}
\end{figure}

The phase structures of two-component BECs with respective VAVSS are
obviously different from those of the usual two-component BECs \cite%
{Ho,Pu,Trippenbach,Adhikari1,Riboli,Navarro,Catelani,Gligoric,LWen,Kuopanportti,Papp,McCarron}
by virtue of the petal structure of the VAVSS. On the other hand, the
equilibrium properties of the present system are also evidently different
from those of DBFM with a bosonic VAVSS \cite{Wen2}. In the latter case, due
to the Pauli exclusion principle, there is no $s$-wave interaction between
identical fermions in the spin polarized state. In addition, for the
separated phases in the latter case, the gap region between two bosonic
petals is always occupied by the Fermi gas, while for the fully separated
phases and inlaid separated phases in the former case there is no particle
occupation in the gap region of two petals of any component due to the
interspecies repulsion and the requirement of symmetric petal structure and
angular momentum conservation. Furthermore, here the inlaid separated phases
and the asymmetric separated phase have no counterparts in DBFM with a
bosonic VAVSS.

\begin{figure}[tbp]
\centerline{\includegraphics*[width=7.6cm]{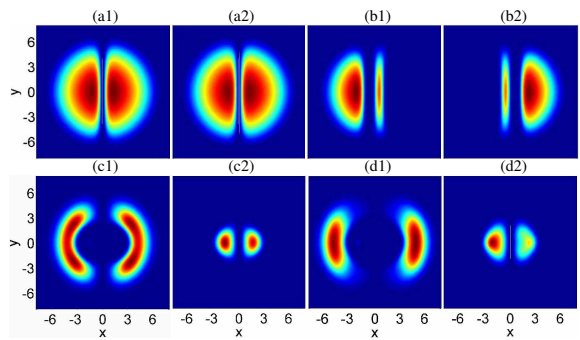}}
\caption{(color online) (a)-(b): Density profiles (a1)-(a2) and (b1)-(b2)
correspond to phases VII and VIII in Fig.1, respectively, where (a) $%
R_{21}=1,R=2$ and (b) $R_{21}=1,R=8$. The other parameters are the same as
those in Fig.1. (c)-(d): Temporal evolution of density profiles, where the
initial state is given by Fig. 2(a) with a random perturbation. The time is
(c) $t=0.4$ and (d) $t=4$. Here $x$ and $y$ are in units of $d_{0}$, and $t$
is units of $1/\protect\omega _{\perp }$.}
\label{Figure4}
\end{figure}

The VAVSS is a collective excitation (a metastable state) of BECs, which is
similar to the case of a single pure vortex state. Although the energy of a
VAVSS is higher than that of a pure vortex \cite{Liu}, the configurations of
two BECs with VAVSS may be long-lived. To verify this point, we perform a
nonlinear stability analysis by monitoring numerically the evolution of a
perturbed stationary solution in Eqs.(\ref{PolarTimeInd1}) and (\ref%
{PolarTimeInd2}). Shown in Figs. 3(c1)-3(d2) are the time evolution of
density profiles, where the initial state is fully separated phase I [see
Fig. 2(a)] with a random perturbation. The system will collapse when the
initially density profiles begin to deform severely. This method can
effectively estimate the lifetime of a VAVSS or a phase structure of the
system. As a matter of fact, the nonlinear stability analysis is widely used
in the study of solitons \cite{Zhang2}. Our simulation shows that the
lifetimes of phases I-VII can reach the order of $100$ ms and the lifetime
of asymmetric separated phase VIII is a bit shorter due to the symmetry
breaking, which allows to be detected in experiments. It is deserved to
mention that similar phase structures also exist in two-component BECs with
VAVSS with higher winding numbers $l_{1k}=l_{1p}=l_{2k}=l_{2p}$. In that
case, the stability of phase structures become lower because of the higher
energy resulted from the higher winding numbers.

\section{Structure of two-component BECs with respective VAVSS with unequal
winding numbers of vortex and antivortex}

In Fig. 4, we present the typical structures of two-component BECs with
respective VAVSS with $l_{jk}=1$ and $l_{jp}=2$ ($j=1,2$). Depending on the
intra- and interspecies coupling strengths, the system shows different
structures such as fully separated phases, inlaid separated phases, and
partially mixed phases, which is similar to the case of $l_{jk}=l_{jp}=1$ to
some degrees. The density profile of each species in a fully mixed phase is
similar to Fig. 4(e2) or Fig. 4(f1). Here an interesting characteristic is
that the density profiles, except those of the inner species in partially
mixed phases and the two species in a fully mixed phase, generally form
closed (highly modulated petal-like) structures. The corresponding phase
profiles are given in Fig. 5, where the value of the phase varies
continuously from $-\pi $ to $\pi $, and the end point of the boundary
between a $\pi $ phase line and a $-\pi $ phase line represents a phase
defect (anticlockwise rotation denotes a vortex while clockwise rotation
denotes an antivortex).

\begin{figure}[tbp]
\centerline{\includegraphics*[width=7.6cm]{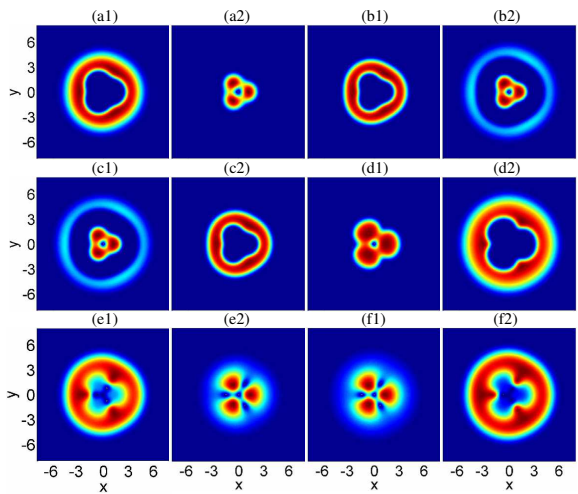}}
\caption{(color online) Density profiles of two-component BECs with
respective VAVSS with $l_{1k}=l_{2k}=1$ and $l_{1p}=l_{2p}=2$. (a) $%
R_{21}=0.2,R=8,$(b) $R_{21}=0.9,R=8,$(c) $R_{21}=1.1,R=7,$(d) $%
R_{21}=2.5,R=5,$(e) $R_{21}=0.6,R=1,$and (f) $R_{21}=1.4,R=1.6$. The other
parameters are the same as those in Fig.2. Here $x$ and $y$ are in units of $%
d_{0}$.}
\end{figure}
\ 

For the cases of fully separated phases, we can see that there is a visible
density hole in the density profile of the interior component [Figs. 4(a2)
and 4(d1)] and a phase defect at the center of the corresponding phase
profile [Figs. 5(a2) and 5(d1)]. The visible density hole is referred to as
a visible vortex (anticlockwise rotation) \cite{Wen3} because the
topological defect is visible in the\textit{\ in situ} density profile and
it contributes to the angular momentum and the energy of the system. At the
same time, there are three phase singularities that locate on the outskirts
of the interior component and form a triangular lattice. Since these phase
defects are invisible in the \textit{in situ} density profile and contribute
to neither the angular momentum nor the energy of the system, they are known
as ghost vortices (exactly speaking, they are ghost antivortices because of
their clockwise rotation) \cite{Wen3,Tsubota,Wen4}. Of particular interest
is the large density hole in the outer component [Figs. 4(a1) and 4(d2)].
From its phase distribution [Figs. 5(a1) and 5(d2)], we can see there are
four phase defects in the region of density hole. The central phase defect
is a vortex (anticlockwise rotation) and the peripheral three phase defects
are antivortices (clockwise rotation) constituting a triangular lattice. The
four topological defects are referred to as hidden vortices \cite%
{Wen3,Wen4,Brtka} because they carry significant angular momentum, though
they are invisible in the \textit{in situ}\ density distribution. Only after
including the hidden vortices can the well-known Feynman rule \cite%
{Wen3,Fetter} be satisfied. The above analysis is also applicable to the
cases of other states such as the inlaid separated-phase states and the
mixed-phase states. For instance, for the partially mixed phases [Figs.
4(e), 4(f), 5(e), and 5(f)], there are four visible vortices in the density
profile of each component in which one is a vortex lying on the trap center
and the other are three antivortices locating on the periphery in a\ shape
of triangular lattice.

\begin{figure}[tbp]
\centerline{\includegraphics*[width=8.4cm]{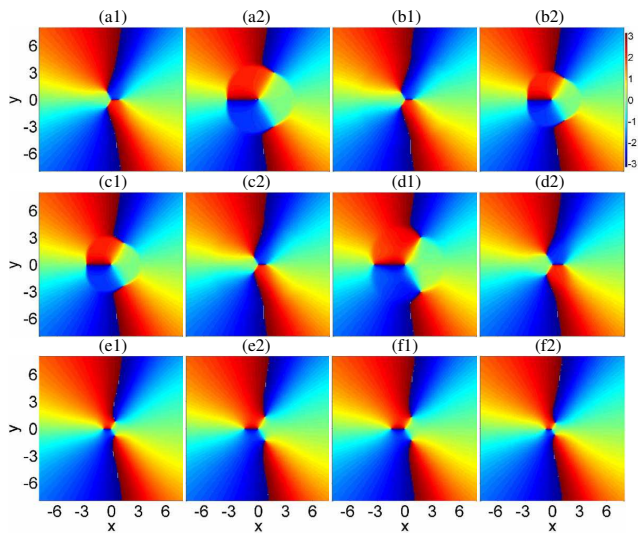}}
\caption{(color online) Phase distributions corresponding to those of the
density profiles of Fig. 5. The value of the phase varies continuously from $%
-\protect\pi $ to $\protect\pi $. The dark color area indicates the lower
phase. Here $x$ and $y$ are in units of $d_{0}$.}
\end{figure}

In general, the phase defects distribute regularly. The trap center is
always occupied by a vortex in both species 1 and species 2. For the
partially mixed phases and the fully mixed phase, the outer antivortices in
both components arrange in a triangular lattice at the almost same azimuth
despite of different radiuses [Figs. 5(e1)-5(f2)]. For the separated phases,
the outer three antivortices in species 1 and those in species 2 arrange
successively along the azimuth direction due to the strong intra- and
interspecies repulsions as shown in Figs. 5(a1)-5(d2).

As for the other cases of $l_{1k}=l_{2k}\neq l_{1p}=l_{2p}$, our simulation
shows that there exist similar phase structures. To understand the formation
mechanism of the above phase structures, we consider a single-component BEC
with a VAVSS with arbitrary unequal winding numbers $l_{k}\neq l_{p}$. As a
general rule, we find that the steady state has $N=l_{k}+l_{p}$ peaks and $%
N+1$ singularities in the phase distribution. When $l_{k}>l_{p}$, there are $%
N$ singularities with unit positive topological charge on the periphery and $%
l_{p}$ singularities with unit negative charge at the center, so that the
total topological charge is $s=N-l_{p}=l_{k}$. When $l_{k}<l_{p}$, there are
$N$ singularities with unit negative charge on the periphery and $l_{k}$
singularities with unit positive charge at the center, so that the total
topological charge is $s=N-l_{k}=l_{p}$. For the case of $l_{k}=l_{p}$ and $%
\alpha ^{2}=\beta ^{2}=1/2$, there is no phase singularity in the phase
distribution due to the zero velocity field \cite{Wen1}. If $l_{k}=l_{p}$
and $\alpha ^{2}>\beta ^{2}$ (or $\alpha ^{2}<\beta ^{2}$), there would be $%
l_{k}$ single-quantum vortices (or antivortices) at the trap center as a
result of the larger vortex (antivortex) ratio $\alpha $ ($\beta $).

For a nonzero-temperature BEC in the presence of thermal atoms, a vortex
state is generally unstable due to the incoherent interactions between the
BEC and the thermal atoms \cite{Rokhsar}. For a zero-temperature BEC, a
singly quantized vortex with repulsive interparticle interaction is stable
while the stability of a multiply quantized vortex is determined by the
property and strength of the interaction \cite{Pu2,Fetter}. Here an
interesting question is what state a VAVSS with $l_{k}=1$, $l_{p}=2,$ and $%
\alpha ^{2}=\beta ^{2}=1/2$ will develop into. Counterintuitively, we find
that the actual state is a vortex-antivortex cluster state consisting of a
singly quantized vortex and three singly quantized antivortices. The density
profile and the corresponding phase profile are similar to Fig. 4(e2) and
Fig. 5(e2), respectively. The underlying physics is that in the presence of
a single-quantum vortex the two-quantum antivortex only decays into three
instead of two singly quantized antivortices because the triangular lattice
\cite{Fetter,Abo-Shaeer} has the lowest energy and is the most stable. It is
well known that the angular momentum of a single vortex depends on its
winding number while the average angular momentum per atom of a condensate
is determined by not only the winding number but also the position of the
vortex in the condensate. Combining with the conservation requirement of the
average angular momentum per atom, then the vortex and the three
antivortices rearrange themselves to a stable spacial structure with the
lowest energy in which the vortex locates on the trap center and the three
antivortices distribute on the outskirts by means of a triangular lattice
form. The similar analysis can be generalized to the other cases. In Ref.
\cite{Pu2}, it is pointed out that a doubly quantized vortex may be stable
in a BEC for certain regions of the interparticle interaction strength. In
the present system, we do not observe similar doubly quantized antivortex in
the equilibrium structures (Fig. 4 and Fig. 5) even we change the values of
the intra- and interspecies interaction strengths in a large scope. The
dynamic processes and details of how the VAVSS decay into the
vortex-antivortex cluster states require further investigation.
Incidentally, when $l_{1k}=l_{1p}\neq l_{2k}=l_{2p}$ our simulation shows
that the equilibrium density profiles may suffer from distortion even
deletion of petals especially for those in a separated phase due to the
symmetry breaking of two group winding numbers and the intra- and
interspecies repulsions, which implies that the phase structures in this
case will become quite complex. Further work would be necessary in order to
understand the more complex structures in this case.

The petal structure of a VAVSS termed initially in Ref. \cite{Kapale}\
remind us of an azimuthon which is proposed lately in nonlinear optics \cite%
{Desyatnikov} and extended recently to BECs \cite{Lashkin}. We show that
there exist evident difference and certain relation between the VAVSS and
the azimuthon. The difference is that our theoretical model and the relevant
studies \cite{Kapale,Liu,Simula,Thanvanthri,Wen1,Wen2} including the
original literature \cite{Kapale} on the VAVSS work in the laboratory frame
while the investigations concerning the azimuthon work in the rotation frame
\cite{Desyatnikov,Lashkin}. Consequently, for the latter case there is an
additional term with respect to angular velocity in the stationary equation
of system \cite{Lashkin}. The relation is that the azimuthon is a special
example of the VAVSS in view of their solution expressions. For the same
winding numbers of vortex and antivortex, the wave function of the VAVSS is
given by $\varphi (r,\theta )\sim \alpha e^{il\theta }+\beta e^{i\delta
}e^{-il\theta }$ [see also Eq. (5) in \cite{Wen1}]. If we take the relative
phase $\delta =0$ and introduce a parameter $p=(\alpha -\beta )/(\alpha
+\beta )$, we can rewrite the wave function as $\varphi (r,\theta )\sim \cos
(l\theta )+ip\sin (l\theta )$. Here the coefficient $\alpha +\beta $ has
been absorbed in the normalization constant of the wave function, and the
parameter $p$ should satisfy the two conditions of $0\leq \left\vert
p\right\vert \leq 1$ and $\alpha ^{2}+\beta ^{2}=1$. In this context, the
azimuthon expressed by $\Phi \sim \cos (l\theta )+ip\sin (l\theta )$ in
Refs. \cite{Desyatnikov,Lashkin} is indeed a particular case of the VAVSS.
Obviously, a VAVSS is not equivalent to an azimuthon when $\delta \neq 0$.
The similar analysis is also applicable to the cases of different winding
numbers of vortex and antivortex.

\section{Conclusion}

We have numerically studied the exact 2D steady structures of two-component
BECs with respective VAVSS. Depending on the winding numbers of vortex and
antivortex and the intra- and intercomponent interactions, the system shows
rich phase configurations such as fully separated phases, inlaid separated
phases, asymmetric separated phase, and partially mixed phases with
(deformed) petal-like component density distributions. For given parameters,
we display a phase diagram of two-components with respective VAVSS with $%
l_{1k}=l_{1p}=1$ and $l_{2k}=l_{2p}=1$. We show that the kinetic energy
plays a key role in determining the structure of the two-component BECs with
respective VAVSS, where the conventional phase separation criterion is
inapplicable and the TF approximation may fail due to the VAVSS, especially
for the cases of VAVSS with large winding numbers. In addition, the
conventional criterion for phase separation and the TF approximation can not
discriminate different separated phases. According to the nonlinear
stability analysis, the typical structures of the system are long-lived,
which allows to be detected and tested in current experiments. The similar
phase structures also exist in the two-component BECs with respective VAVSS
with unequal winding numbers of vortex and antivortex. In this case, the
density profile of each species forms a closed (highly modulated petal-like)
structure except the fully mixed phase and the inner species in a partially
mixed phase. In particular, an interesting vortex-antivortex cluster state
occurs in each component in any of the possible phase structures, where the
vortices and antivortices appear in the form of visible vortex, or hidden
vortex, or ghost vortex. Furthermore, a general relation between the
vortex-antivortex cluster states and the winding numbers of vortex and
antivortex is revealed and analyzed. Finally, we show that the azimuthon
proposed recently in nonlinear optics and in BECs is a special example of
VAVSS in terms of their solution expressions.

\begin{acknowledgments}
L.W. thanks Chuanwei Zhang, Biao Wu and Yongping Zhang for helpful
discussion. This work was supported by the NSFC under Grants No. 11047033
and No. 10847143, the International Cooperation Program by Shandong
Provincial Education Department, and the NSF of Shandong Province under
Grant No. Q2007A01.
\end{acknowledgments}

\end{document}